\documentclass[fleqn,usenatbib]{mnras}

\usepackage{newtxtext,newtxmath}

\usepackage[T1]{fontenc}

\DeclareRobustCommand{\VAN}[3]{#2}
\let\VANthebibliography\thebibliography
\def\thebibliography{\DeclareRobustCommand{\VAN}[3]{##3}\VANthebibliography}

\usepackage{graphicx}	
\usepackage{amsmath}	
\usepackage{orcidlink}
\usepackage{float}

\usepackage[normalem]{ulem} 

\title[B fields and satellite stripping]{The role of magnetic fields in ram pressure stripping of satellite galaxies in the circumgalactic medium around massive galaxies}

\author[T. A. Rintoul et al.]{
Thomas A. Rintoul$^1$\thanks{E-mail: RintoulTA@cardiff.ac.uk}\orcidlink{0009-0006-9774-0807},
Freeke van~de~Voort$^1$
\orcidlink{0000-0002-6301-638X},
Andrew T. Hannington$^1$
\orcidlink{0009-0006-4444-2566},
R\"udiger Pakmor$^2$
\orcidlink{0000-0003-3308-2420},
\newauthor{
Rebekka Bieri$^3$
\orcidlink{0000-0002-4554-4488},
Maria Werhahn$^2$
\orcidlink{0000-0003-4984-4389},
and
Rosie Y. Talbot$^2$
\orcidlink{0000-0001-9393-7879}
}
\vspace*{0.1cm}\\
$^{1}$Cardiff Hub for Astrophysics Research and Technology, School of Physics and Astronomy, Cardiff University, Queen’s Buildings, Cardiff CF24 3AA, UK
\\
$^{2}$Max-Planck-Institut f\"{u}r Astrophysik, Karl-Schwarzschild-Str. 1, D-85748, Garching, Germany
\\
$^{3}$Department of Astrophysics, University of Zurich, Zurich, Switzerland
\\
}

\date{Accepted --. Received 2025 June 23; in original form 2025 June 23}

\pubyear{\the\year{}}

\begin{document}


\label{firstpage}
\pagerange{\pageref{firstpage}--\pageref{lastpage}}
\maketitle

\begin{abstract}
The presence of magnetic fields in galaxies and their haloes could have important consequences for satellite galaxies moving through the magnetised circumgalactic medium (CGM) of their host. 
We therefore study the effect of magnetic fields on ram pressure stripping of satellites in the CGM of massive galaxies.
We use cosmological `zoom-in' simulations of three massive galaxy haloes ($M_{\rm{200c}} = 10^{12.5-13}$~M$_\odot$), each simulated with and without magnetic fields.
Across our full sample of satellite galaxies (11 with magnetic fields and 10 without), we find that the fraction of gas retained after infall through the CGM shows no population-wide impact of magnetic fields. 
However, for the most massive satellites, we find that twice as much gas is stripped without magnetic fields.
The evolution of a galaxy's stripped tail is also strongly affected. 
Magnetic fields reduce turbulent mixing, inhibiting the dispersion of metals into the host CGM.
This suppressed mixing greatly reduces condensation from the CGM onto the stripped tail.
By studying the magnetic field structure, we find evidence of magnetic draping and attribute differences in the stripping rate to the draping layer.
Differences in CGM condensation are attributed to magnetic field lines aligned with the tail suppressing turbulent mixing.
We simulate one halo with enhanced resolution in the CGM and show these results are converged with resolution, though the structure of the cool gas in the tail is not.
Our results show that magnetic fields can play an important role in ram pressure stripping in galaxy haloes and should be included in simulations of galaxy formation.

\end{abstract}




\section{Introduction}


The circumgalactic medium (CGM) is the gaseous component of the dark matter-dominated halo around a galaxy.
It contains gas accreted from the intergalactic medium (IGM), as well as gas ejected from the interstellar medium (ISM).
The CGM plays a vital role in galaxy evolution, providing a vast reservoir of gas which can be accreted onto the central galaxy, fuelling star formation within.

The CGM is multiphase, with temperatures ranging from ${\sim}10^{4}$~K up to the virial temperature (${\sim}10^{7}$~K for haloes of $10^{13}$~M$_\odot$). If heated by sources of feedback, gas can even exceed the virial temperature.
While the many phases of the CGM lead to emission across the electromagnetic spectrum (X-ray:~\citealt{bogdan2013, comparat2022, chadayammuri2022}, ultraviolet:~\citealt{meiring2013, henry2015, rubin2014}, optical:~\citealt{rubin2012}, radio:~\citealt{das2020}), its diffuse nature makes this emission very faint.
For this reason, many observations of the CGM focus on absorption by the CGM. Foreground absorption lines in spectra of background sources can probe more diffuse gas in the CGM~\citep{lehner2015, tumlinson2017}.

Hydrodynamic simulations allow us to explore and test the physical processes governing the evolution of galaxies and the CGM.
Cosmological simulations account for the formation history and large-scale environment of galaxies and are able to produce a reasonable galaxy population which broadly reproduces the galaxy population of the real universe~\citep{crain2023}.

Satellite galaxies contribute gas to the CGM of the central galaxy~\citep{faucher-giguere2023, ramesh2024}.
There are several mechanisms by which satellite galaxies lose gas into their surroundings.
These include `internal' processes such as outflows due to stellar and active galactic nucleus (AGN) feedback.
Environmental effects also play a significant role.
An important mechanism by which satellites lose gas to their host is through ram pressure stripping~\citep{gunn1972}, i.e. the removal of gas due to ram pressure from the medium the satellite moves through.

The removal of a galaxy's gaseous halo by ram pressure stripping can substantially suppress the rate of gas accretion onto the galaxy.
This results in a decline in the star formation rate - a process known as `starvation' or `strangulation'~\citep{larson1980, balogh2000}.
This can lead to the long-term quenching of star formation, particularly in satellite galaxies~\citep{vandevoort2017, walters2022}.

Ram pressure stripping has been studied extensively in cluster environments ($M_{\rm{200c}} \gtrsim 10^{14}$~M$_\odot$).
Simulations and observations both show evidence of ram pressure stripped cluster galaxies with long tails of dense stripped gas, often referred to as `jellyfish galaxies'~\citep{yun2019, muller2021, rohr2023a, sparre2024}.
Some of these tails are known to host in-situ star formation~\citep{boselli2005}.
The Intracluster Medium (ICM) is hotter than the CGM of a massive galaxy by a factor of 10-100.
It is also more relaxed than the CGM because it is less affected by the feedback processes of a single galaxy~\citep{donahue2022, lepore2025}.
The ICM has a similar, though somewhat higher density to the CGM around massive galaxies, but the velocities of satellite galaxies in cluster environments are typically substantially higher~\citep{yun2019}.
Ram pressure $\propto \rho v^2$, where $\rho$ is the density of the medium and $v$ is the velocity of the galaxy relative to the medium.
Because they tend to orbit faster, cluster satellites typically experience much greater ram pressure than their counterparts in lower mass haloes.

Wind-tunnel simulations show that ram pressure can efficiently strip a satellite galaxy over a few hundred Myrs.
Some recent wind-tunnel simulations of ram pressure stripping implement a uniform but time-dependent wind to better model the ram pressure experienced by a satellite orbiting in the ICM~\citep{tonnesen2019, sparre2024}.
A time-dependent wind prevents over-stripping of the satellite as seen in some simulations with constant winds~\citep{tonnesen2019}.
This has provided a more realistic yet controlled environment in which to explore the effects of ram pressure stripping.
Works by \cite{zhu2024} and \cite{ghosh2024} find that the CGM of satellites is easily stripped, though the ISM may still be retained over long periods provided that the mass and therefore the gravitational restoring force is large enough.

Using cosmological simulations, \cite{yun2019} found that jellyfish galaxies are ubiquitous throughout the ICM.
Cluster jellyfish galaxies experience strong ram pressure stripping, resulting in a long tail of dense gas.
\cite{rohr2023a} found that in cluster environments, ram pressure stripping is capable of completely stripping and quenching a galaxy in a single orbit (infall to apocentre) and that half of the cluster jellyfish galaxies were quenched by ram pressure stripping at $z=0$.
In lower mass systems, satellites tend to have lower velocities relative to the ambient medium and therefore, experience less ram pressure.
\cite{simpson2018} investigated quenching of satellite galaxies due to ram pressure stripping in Milky Way-mass haloes. 
They found that the majority of satellites with $M_{\rm{stellar}} < 10^8 M_{\odot}$ within $300$~kpc of the central galaxy are quenched by ram pressure stripping at $z=0$. More massive satellites ($M_{\rm{stellar}} \gtrsim 10^9$~M$_\odot$) are typically not (yet) quenched at $z=0$.
Understanding the impact of ram pressure stripping is therefore important for understanding the evolution of satellite galaxies.

Magnetic fields have been shown to have potentially significant effects on the overall evolution of galaxies~\citep{whittingham2021} and the CGM~\citep{vandevoort2021}.
These include systematically larger galaxy discs and better retention of gas following mergers~\citep{whittingham2021}, as well as affecting the structure and mixing of the CGM and the propagation of outflows~\citep{vandevoort2021}.
Note, however, that these effects are not seen in all simulations. \cite{hopkins2020} find no strong effects from magnetic fields in the FIRE-2 simulations.
\cite{sparre2020, sparre2024a, sparre2024} conducted a series of wind-tunnel simulations exploring the effects of ICM winds on gas clouds and galaxies. 
\cite{sparre2020} found that cool gas clouds survive longer in a hot wind featuring a turbulent magnetic field compared to a wind with a uniform field, or no magnetic field.
Similarly, \cite{sparre2024} found that when cluster-like magnetic fields (i.e. turbulent) are included, galaxies also experience less ram pressure stripping. 

While there have been a considerable number of investigations into ram pressure stripping in cluster environments, fewer studies focus on its effect in galaxy groups or around massive galaxies.
In this work, we use cosmological `zoom-in' simulations of 3 massive haloes~($M_{\rm{200c}} = 10^{12.5}-10^{13}$~M$_\odot$) which host a number of satellite galaxies.
These simulations form part of the Simulating the Universe with Refined Galaxy Environments (SURGE) project.
We simulate these haloes with and without magnetic fields to investigate the effect of magnetic fields on ram pressure stripping of satellite galaxies.
We show that while there is not a clear population-wide impact, ram pressure stripping of massive, gas-rich satellites, and the evolution of the subsequent stripped tail is substantially altered by the presence of magnetic fields.
The simulation method is described in Section~\ref{sec:methods}.
In Section~\ref{sec:results}, we present our results on the effect of magnetic fields on ram pressure stripping of satellites.
We explore the effects of additional resolution in the CGM in Section~\ref{sec:CGM-Refinement}.
Our findings are summarised and discussed in Section~\ref{sec:disc_and_conc}.
All length scales are given in proper coordinates.

\section{Methods}\label{sec:methods}

\begin{table}
	\centering
	\caption{Properties of the host haloes at $z=0$: name of the halo, inclusion of $B$ field, total halo mass ($M_{\textrm{200c}}$) and its virial radius ($R_{\textrm{200c}}$), as well as the imposed spatial resolution limit from CGM refinement, if applicable.}
	\label{tab:halo_table}
	\begin{tabular}{lcccc}
		\hline
		Halo          & $B$   & ${\log_{10}} M_{\textrm{200c}}$ & $R_{\textrm{200c}}$ & CGM        \\
        Name          & Field & [M$_\odot$]                     & [kpc]               & Refinement \\
		\hline
		h8\_std       & Yes   & 13.01                           & 458.7               & --         \\
		h8\_noB       & No    & 13.03                           & 462.4               & --         \\
		h8\_1kpc      & Yes   & 13.00                           & 455.3               & {+}1 kpc   \\
		h8\_noB\_1kpc & No    & 13.03                           & 462.8               & {+}1 kpc   \\
		h7\_std       & Yes   & 13.07                           & 478.1               & --         \\
		h7\_noB       & No    & 13.10                           & 487.7               & --         \\
		h3\_std       & Yes   & 12.51                           & 310.8               & --         \\
		h3\_noB       & No    & 12.52                           & 314.1               & --         \\
		\hline
	\end{tabular}
\end{table}

\begin{figure*}
    \includegraphics[width=\textwidth]{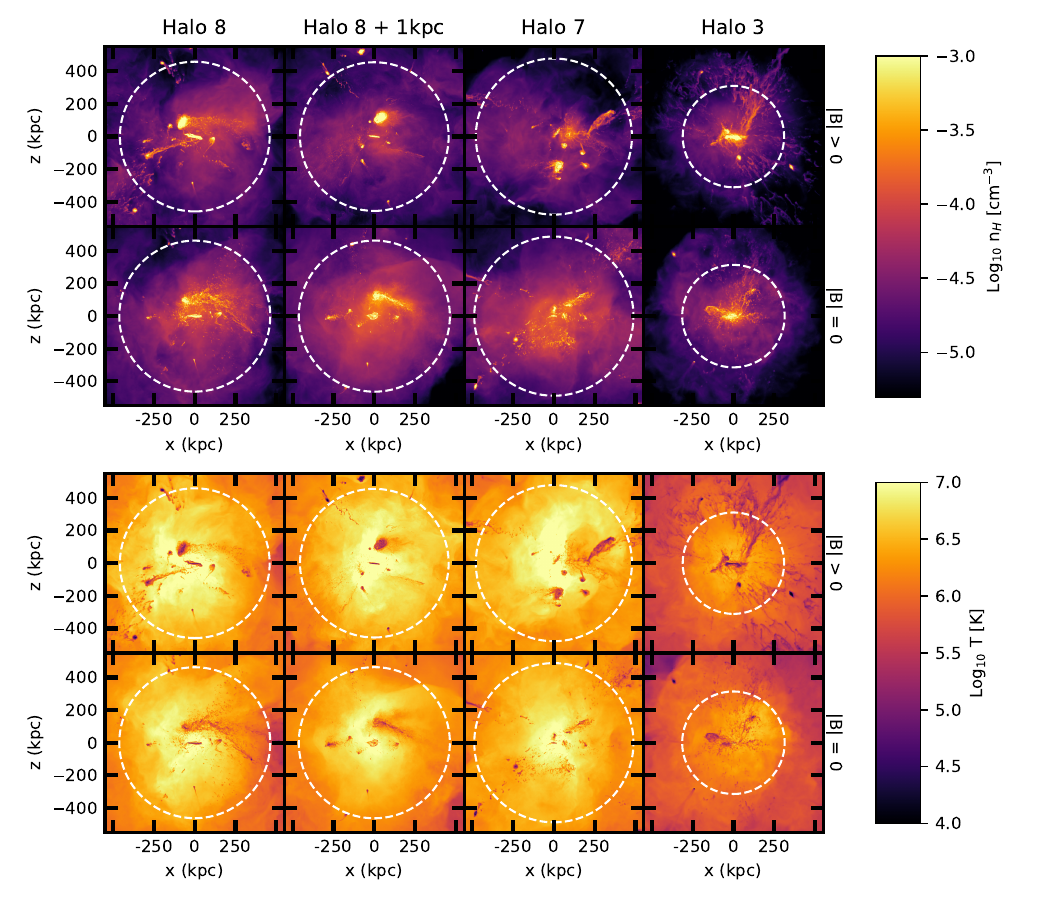}
    \caption{Edge-on projections of the hydrogen number density (top) and mass-weighted temperature (bottom) for 3 different galaxy haloes at $z=0$. We assume solar metallicity, $Z_\odot = 0.0127$.
    All haloes are rotated so the central galaxy is displayed `edge-on'.
    All haloes are projected at a depth of 200 kpc.
    Haloes 7 and 8 have a halo mass $M_{200c} \approx 10^{13}$~M$_\odot$, while Halo 3 has a halo mass of approximately $10^{12.5}$~M$_\odot$. See Table~\ref{tab:halo_table} for full details. The simulations in the top row of each group of panels include magnetic fields, whereas those in the the bottom row do not. Halo 8 was resimulated with 1 kpc spatial refinement in addition to the existing mass-refinement target. The virial radius, $R_{\rm{200c}}$, for each halo is indicated by a dashed white circle.
    Haloes without magnetic fields generally have higher gas densities in the CGM, particularly in the vicinity of the central galaxy. Temperatures are slightly higher in outer regions with magnetic fields but are otherwise similar. Given that Halo 3 has a lower halo mass, its virial radius and CGM temperature are lower than for the more massive haloes. Additionally, the central galaxy gas discs are smaller without magnetic fields and satellite galaxy tails appear denser and broader.}
    \label{fig:all_haloes_nh-t}
\end{figure*}

\begin{figure*}
    \includegraphics[width=\textwidth]{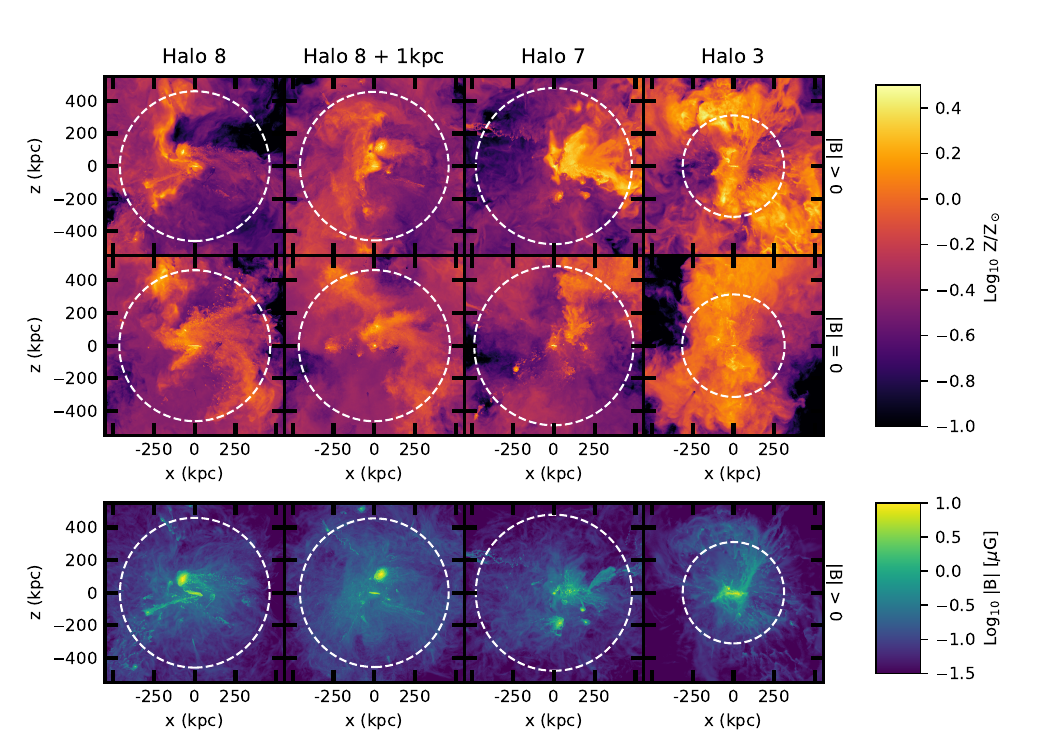}
    \caption{Same as Fig.~\ref{fig:all_haloes_nh-t} but for metallicity (top) and magnetic field strength (bottom), both mass-weighted.
    Metals are more smoothly distributed in simulations without magnetic fields. The magnetic field strengths are similar in all haloes. Higher magnetic field strengths are found in the central and satellite galaxies.
    The cool gas in satellite tails has lower field strengths than the gas in galaxies, but higher than the ambient hot halo gas.
    In Halo 3, we see a region of high magnetic field strength extending from the galaxy to the virial radius. This is co-spatial with cool, metal-rich gas expelled from the galaxy disc. Therefore, it is clear that not all regions of higher magnetic field strength in the CGM are due to ram pressure stripping of satellites.}
    \label{fig:all_haloes_z-b}
\end{figure*}

We conducted cosmological zoom-in simulations of 3 haloes with $z=0$ halo masses ($M_{\textrm{200c}}$) in the range $10^{12.5}$ to $10^{13}\,\mathrm{M_\odot}$.
Here, we define $M_{\textrm{200c}}$ and $R_{\textrm{200c}}$ such that the volume enclosed by a sphere of radius $R_{\textrm{200c}}$ has an average density 200 times the critical density of the universe, and $M_{\textrm{200c}}$ is the mass enclosed within $R_{\textrm{200c}}$.

Haloes were selected from the dark matter-only EAGLE simulation box~\citep{schaye2015} with a co-moving box size of 100 Mpc.
The haloes were re-simulated at significantly higher resolution with a zoom-in region centred on the selected halo, extending to a radius of $5\times R_{\textrm{200c}}$.
Details of all haloes are listed in Table~\ref{tab:halo_table}.
These simulations form part of the SURGE project (see also \citealt{vandevoort2019}).
Some of the simulations presented here are also discussed in \cite{pakmor2024} and \cite{werhahn2025} in the context of understanding the amplification of magnetic fields.

We simulated each halo with and without magnetic fields.
In simulations with magnetic fields, we seed a uniform magnetic field at $z = 127$ with a co-moving strength of $10^{-14}$~G.
The choice of a uniform seed field rather than a more physically motivated seed field is unlikely to matter at $z \le 1.5$ as a turbulent magnetic dynamo amplifies the field to saturation before this time~\citep{pakmor2014, marinacci2015, garaldi2021, pakmor2024}.
For simulations without magnetic fields, we set the magnetic field seed value to zero and, therefore, the magnetic field remains exactly zero for the whole simulation time. The simulations are otherwise identical.

Our simulations were run using the \textsc{arepo} code \citep{springel2010, pakmor2016, weinberger2020}.
\textsc{arepo} solves the equations of ideal magnetohydrodynamics (MHD) with a finite volume second order Runge-Kutta integration scheme~\citep{pakmor2016} on a moving, unstructured Voronoi mesh using the \cite{powell1999} approach for divergence control.
Our simulations adaptively refine gas cells by imposing a target cell mass $M_{\textrm{cell}} = 5\times 10^4 M_\odot$.
Cells are refined if their mass exceeds $2\times m_{\rm{target}}$ and de-refined if their mass drops below $0.5\times m_{\rm{target}}$.

In addition to our mass-refined simulations, we re-simulate one $10^{13}$~M$_\odot$ halo (Halo 8) with additional spatial refinement (see \citealt{vandevoort2019}) which limits the volume of gas cells in the CGM to $V_{\rm{target}} \approx 1$ kpc$^3$.
Similar to the mass target, cells are refined if their volume exceeds $2\times V_{\rm{target}}$ and de-refined if their volume dropped below $0.5\times V_{\rm{target}}$.
This volume refinement criterion extends to $1.2 \times R_{\rm{200c}}$ from the centre of any halo with $M_{\rm{200c}} \ge 5\times10^8$~M$_\odot$.
This achieves a near-uniform spatial resolution in the CGM.

Cell refinement methods in \textsc{arepo} make it impossible (over cosmologically relevant time periods) to follow individual gas cells.
To overcome this, we use Monte-Carlo tracer particles which move stochastically to follow mass when exchanged between resolution elements representing various baryonic mass components - gas cells, star and wind particles, and black holes~\citep{genel2013, defelippis2017, grand2019}.
This allows us to follow mass fluxes through the simulation. 
At the start of the simulation, one tracer particle is initialised in every gas cell.
Over time, tracer particles will move into new cells.
Individual gas cells can hold multiple tracer particles and it is not guaranteed that every cell will contain a tracer.

We use the Auriga galaxy formation model which is described fully in \cite{grand2017} and is summarised here.
We include radiative cooling of hydrogen and helium, metal line cooling and heating from a spatially uniform and time-evolving UV background radiation field \citep{vogelsberger2013} with photoionisation rates based on the UVB intensity of \cite{faucher-giguere2009}, including corrections for self-shielding~\citep{rahmati2013}.
The ISM is modelled with a subgrid model, as described in \cite{springel2003}, featuring a stochastic model for star formation (with threshold density of $n_{\rm{H}}\approx0.11$~cm$^{-3}$), metal production and stellar feedback-driven winds~\citep{vogelsberger2013}.
Additionally, we include a model for the formation and growth of supermassive black holes and associated feedback~\citep{grand2017}.
We use cosmological parameters from Planck Collaboration XVI (\citeyear{planckcollaboration2014}) of $\Omega_{m} = 0.307$, $\Omega_{b} = 0.048$, and $\Omega_{\Lambda} = 0.693$ with $H_{0} = 100 h$ km s$^{-1}$ Mpc$^{-1}$ where $h = 0.6777$.
The Auriga model has been extensively used to simulate Milky Way-mass galaxies~\citep{grand2017, grand2021}.

Galaxies and their host haloes are identified using a friends-of-friends (FoF) algorithm~\citep{davis1985}.
Stellar, gas and black hole particles are assigned to the same FoF-halo as their nearest dark matter particle.
Subhaloes within FoF-haloes are identified using the \textsc{subfind} algorithm \citep{springel2001, dolag2009} which defines `subhaloes' as gravitationally-bound over-densities accounting for both dark matter and baryons.
The subhalo with the lowest gravitational potential energy is defined as the main halo containing the central galaxy, with other subhaloes described as satellites.
Baryonic resolution elements (gas cells, star particles, black hole particles) which are gravitationally bound to the subhalo are associated with that subhalo and contribute to its recorded mass.
To trace haloes over cosmologically relevant time periods, merger trees are produced in a post-processing step using the code developed by~\cite{springel2005}. 
This links the individual objects catalogued by \textsc{subfind} within an individual output and across simulation outputs.

This work focuses on the effect of magnetic fields on satellite stripping.
Not all subhaloes within a simulation host luminous satellites.
Additionally, not all luminous satellites are suitable for exploration in this work.
Satellites must be sufficiently massive such that they are resolved adequately.
The key parameter in this work is the inclusion of magnetic fields, therefore it is vital that the magnetic fields are converged during the period of interest for these satellites.
As discussed in~\cite{pakmor2024}, the velocity field in the ISM is highly turbulent on small scales which drives a small-scale turbulent dynamo capable of amplifying the magnetic field strength in the galaxy.
This amplification continues until the dynamo is saturated, typically when the magnetic energy density reaches a few 10 per cent of the turbulent energy density.
The saturation time of the magnetic field depends on resolution.
If the resolution is too low, the field strength will not converge until almost $z=0$.
Given our baryonic mass resolution of $5\times10^{4}$~M$_\odot$ and the results of~\cite{pakmor2024}, these considerations limit us to satellite galaxies with pre-stripping mass $M_{\rm{total, init}}\ge~3\times~10^{10}$~M$_\odot$ at $z\le0.5$.
Satellites above this mass limit will have sufficiently amplified and converged magnetic fields and will have developed an azimuthal magnetic field structure in the disc.

To follow the evolution of the satellite while undergoing ram pressure stripping, we only include satellites in our sample which are  present within the host CGM for an extended period of time.
Finally, we wish to only examine galaxies on first infall, because satellites which have previously orbited within the virial radius will have significantly stronger internal magnetic fields as discussed in~\cite{werhahn2025}.
Therefore, we exclude any satellite galaxies which have previously completed a full orbit.

To that end, we introduced the following selection criteria:
\begin{enumerate}
    \item Satellites must be located within $R_{\rm{200c}}$ for at least 5 simulation outputs (${\sim}500$ Myr) between $z=0.5$ and $z=0$ to allow their evolution to be followed;
    \item Satellites must have a total mass of $M_{\rm{total, init}}\ge~3\times~10^{10}$~M$_\odot$ (including gas, stars, dark matter and black holes) before being stripped, and;
    \item Satellites must be on their first infall.
\end{enumerate}
This results in a sample of 11 satellite galaxies from the 3 simulations with magnetic fields and 10 satellites from those without magnetic fields.

\section{Results}\label{sec:results}

\subsection{An overview of the satellite galaxy sample}

To understand how ram pressure stripping affects our simulated satellites, we first aim to understand the CGM environment through which they move.
To this end, Figures~\ref{fig:all_haloes_nh-t} and~\ref{fig:all_haloes_z-b} show projections of the hydrogen number density, temperature, metallicity, and magnetic field strength (if applicable) for our 3 simulated haloes at $z=0$.
Temperature, metallicity and magnetic field strength have all been mass-weighted to focus attention on the cool gas in the CGM.
Magnetic field strength is not typically mass-weighted in other cosmological simulation works. Instead, it is volume-weighted which lends more weight to the hot phase of the CGM.
The cool gas in stripped satellite tails accounts for a significant fraction of the tail mass, but a much smaller fraction of the volume in the wake of satellite galaxies.
Because we focus on the satellites and their tails, we make the choice to use mass-weighted magnetic field strength to gain a better understanding of how magnetic fields affect cool gas in the tail.
Also shown are the re-simulations of Halo~8 with 1~kpc CGM refinement (see Sec.~\ref{sec:CGM-Refinement} for details on the effect of increasing the spatial resolution in the CGM).
The white dashed circles show the virial radius of the halo.
Throughout our analysis, we assume a solar metallicity value, $Z_\odot = 0.0127$.

We see that the central gaseous discs are smaller in the simulations which do not include magnetic fields (as seen in \citealt{whittingham2021} for lower mass galaxies).
In the case of Halo 7, the central galaxy has a very small gas disc both with and without magnetic fields due to prior mergers (most recently at $z=0.17$) removing the majority of the gas.
In simulations without magnetic fields, we see higher (${\sim}0.5$ dex) CGM densities in central regions.
This contrasts results in Milky Way-mass ($M_{\rm{200c}} \approx 10^{12}$ M$_\odot$) haloes which showed lower CGM densities (${\sim}0.5$ dex difference) within the inner 50 kpc in simulations without magnetic fields~\citep{vandevoort2021}.
Outer regions are slightly hotter when magnetic fields are included as in $10^{12}$~M$_\odot$ haloes~\citep{vandevoort2021}.
We do not see any other significant differences in temperature.
Metals are also more smoothly distributed in simulations without magnetic fields as in $10^{12}$~M$_\odot$ haloes~\citep{vandevoort2021}.
The magnetic field strength is generally high for gas in the central and satellite galaxies, with lower field strengths in the CGM.
The magnetic field strength in the cool gas in satellite tails is lower than in gas in galaxies, but higher than the ambient hot halo gas.
We note, however, that we see similar magnetic field strengths associated with cool, metal-rich outflows in Halo 3, indicating that higher field strengths in the CGM are not solely due to ram pressure stripping.
As expected given its lower halo mass, the temperature and extent of the CGM are lower for Halo 3.
With magnetic fields, the density of the CGM in Halo 3 is similar to the more massive haloes.
Without magnetic fields, the density is slightly lower by a factor of ${\sim}3$.
All three haloes host a number of satellite galaxies that are undergoing ram pressure stripping at $z=0$.
Tails extending from these satellite galaxies are somewhat denser with larger clumps without magnetic fields.

\begin{figure}
    \centering
	\includegraphics[width=\columnwidth]{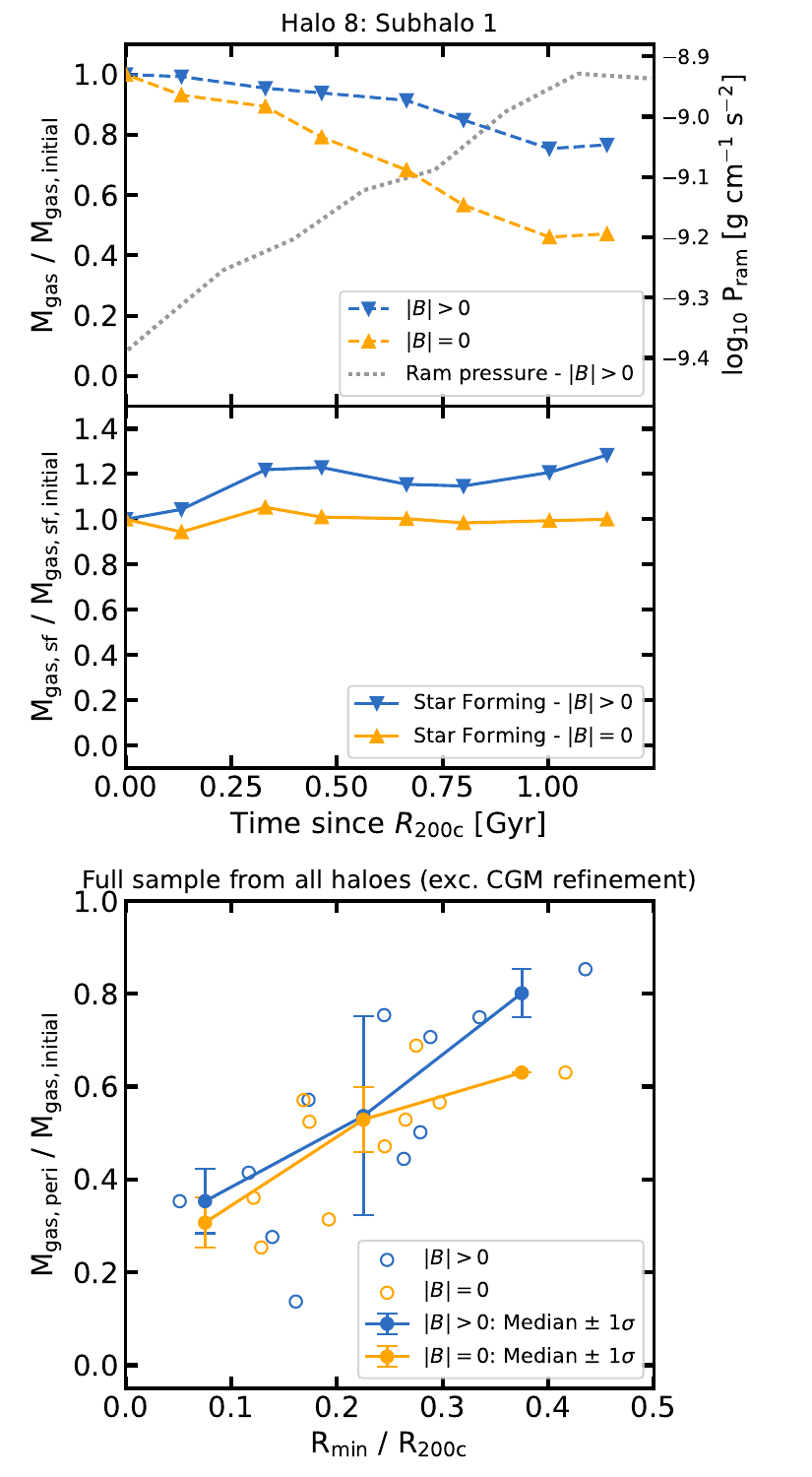}
    \caption{The top panel shows the gas mass evolution of Halo 8: Subhalo~1, with and without magnetic fields, as a function of time since crossing $R_{\rm{200c}}$ of the main halo. It takes approximately 1~Gyr for the satellite to travel between $R_{\rm{200c}}$ and its (approximate) pericentric distance, $R_{\rm{min}}$, where $R$ is the distance between the satellite and the central galaxy. With $B$ fields, ${\sim}20\%$ of total gas mass (ISM+CGM) is lost in this time. Without $B$ fields, ${>}50\%$ of gas is stripped. The dotted grey line shows the ram pressure experienced by the subhalo in the simulation with magnetic fields. The ram pressure is calculated as the product of the mean density ahead of the subhalo, and the square of the subhalo velocity. The evolution of the ram pressure is similar without magnetic fields.
    The middle panel shows that the star-forming gas mass remains reasonably stable in both simulations indicating that most gas is lost from the satellite's CGM, not the ISM.
    The bottom panel shows the retained gas fraction at pericentre as a function of the distance between pericentre and the central galaxy for a sample of satellite galaxies with total pre-stripping mass, $M_{\rm{total, init}} \ge 3 \times 10^{10}$~M$_\odot$, and which meet the other criteria given in Sec.~\ref{sec:methods}. As with Halo~8: Subhalo~1, we follow each satellite from $R_{\rm{200c}}$ to pericentre. As expected, galaxies with smaller pericentric distances are more strongly stripped. Median values suggest stripping is slightly stronger without $B$ fields. However, we see considerable scatter and the two populations are consistent with each other overall.}
    \label{fig:mass_loss_comparison}
\end{figure}

\begin{figure*}
	\includegraphics[width=\textwidth]{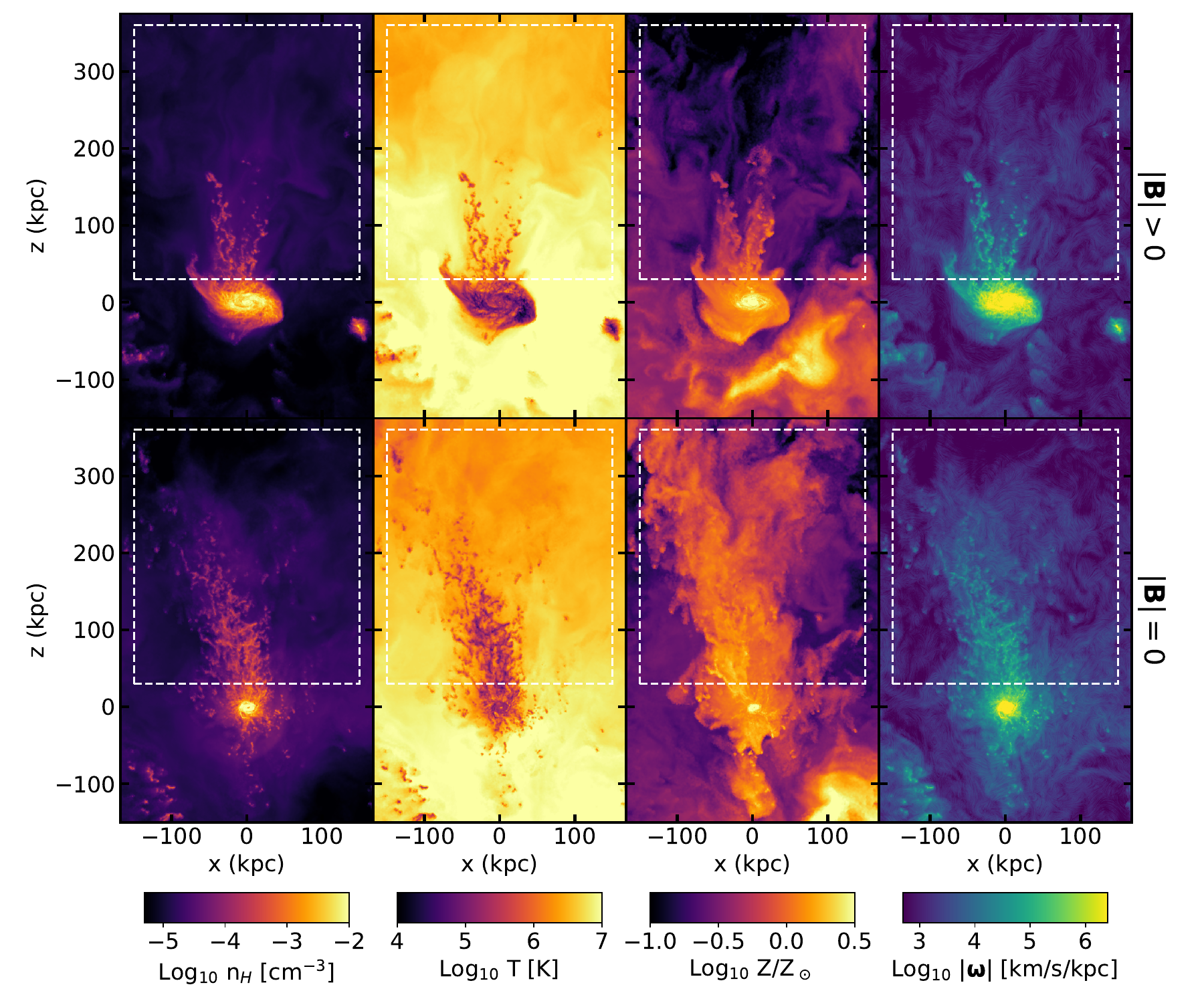}
    \caption{Projections of the hydrogen number density, temperature, metallicity and vorticity for the largest satellite of Halo 8 with (top panel) and without (bottom panel) magnetic fields. 
    We project this satellite with a projection depth of 50 kpc.
    Vorticity field lines are overlaid using the Line Integral Convolution method\protect~\citep{cabral1993}\protect. The galaxy is rotated such that the satellite velocity vector points in the negative $z$-direction. 
    The satellites are shown when they have reached pericentre and thus the moment of maximal stripping for the satellites. 
    The stripped tail is much more extensive and the satellite galaxy disc is smaller in the simulation without magnetic fields. The metallicity and, to a lesser extent, the vorticity, show elevated values in the CGM of the host halo around but outside the tail in the simulation without magnetic fields. This indicates increased turbulence between the ram pressure stripped gas and the host CGM.
    Dashed white rectangles denote the region from which tracer particles are selected in Sec.\protect~\ref{sec:tracer-analysis-results}\protect.
    }
    \label{fig:Satellite_Gas_with_without_B}
\end{figure*}

In Fig.~\ref{fig:mass_loss_comparison}, we explore the overall effects of ram pressure stripping on the gas in our satellite galaxy population.
The top two panels show the evolution of the total and star-forming gas mass of the largest satellite in Halo~8 (hereafter, Halo~8: Subhalo~1) over the course of its infall towards the central galaxy.
This is the largest satellite in our sample so is not representative of the full sample.
We focus our analysis on this satellite but we also find similar results for the other large satellite in our sample (Halo~7: Subhalo~1).
We begin following this satellite when it crosses $R_{\rm{200c}}$ of the host halo and follow its evolution to pericentre.
This satellite has a very elliptical orbit such that within the CGM, the orbit is near parabolic.
It reaches approximate pericentre, at $z\approx 0.01$ at a distance of $R_{\rm{min}} = 112$~kpc from the central galaxy with a velocity of ${\sim}600$~km/s, relative to the central galaxy.
$R_{\rm{min}}$ is the smallest separation between the satellite and the central galaxy in any simulation output.
There is no reason that a simulation output should coincide with a satellite at pericentre.
Due to the time between simulation outputs, any measurements or visualisations may be displaced from true pericentre by ${\sim}100$~Myr.

The first panel of Fig.~\ref{fig:mass_loss_comparison} shows the fraction of total gas retained by the satellite.
The second panel shows the fraction of star-forming gas retained.
The initial total gas mass is taken as the mass of all gas cells associated with the subhalo by \textsc{subfind} in the simulation output immediately prior to the satellite crossing $R_{\rm{200c}}$ of the host halo.
For Halo 8: Subhalo 1, $M_{\rm{gas, initial}}~=~7.9~\times~10^{10}$~M$_{\odot}$ with magnetic fields and $M_{\rm{gas, initial}}~=~6.9~\times~10^{10}$~M$_{\odot}$ without magnetic fields (a difference of ${\approx}10\%$).
The star-forming gas mass is taken as the sum of the mass of all star-forming gas cells associated with the satellite by \textsc{subfind}, where we take the initial mass at the same time as initial total gas mass.
The satellite has $M_{\rm{gas, sf, init}} = 4.5\times10^{10}$~M$_\odot$ and $M_{\rm{gas, sf, init}} = 3.5\times10^{10}$~M$_\odot$, with and without magnetic fields, respectively.

We also show the ram pressure experienced by the satellite during infall in the first panel.
The ram pressure is calculated as the product of the mean gas density ahead of the satellite, and the square of the satellite velocity relative to the host CGM.
The mean gas density was calculated for a cylindrical region ahead of the satellite spanning from 45~kpc to 60~kpc ahead of the satellite, with a radius of 100~kpc from the central axis of the galaxy.
This region was selected such that no cool gas from the satellite disc was included, but that the satellite would traverse the region before the next simulation output.
As expected, ram pressure increases over time as the satellite infalls from the virial radius to pericentre.
In this time, the CGM density increases slightly while the satellite velocity increases significantly, leading to the rise in ram pressure.
We plot ram pressure only for the simulation with magnetic fields. The evolution is very similar for the simulation without magnetic fields.

Over the course of ${\sim}1$~Gyr, we see that Subhalo 1 in Halo 8 retains 77\% and 46\% of its total pre-stripping gas with and without magnetic fields, respectively.
Without magnetic fields, approximately twice as much gas is lost from this satellite.
Some "lost" gas has been converted into stars. 
However, the increase in stellar mass in this time is negligible compared to the decrease in total gas mass.

Despite a significant reduction in total gas mass, the star-forming gas mass in this satellite, both with and without magnetic fields, is fairly constant over the same period.
This indicates that gas is primarily stripped from the more loosely bound CGM of the satellite, rather than the more tightly bound ISM.
In the simulation with magnetic fields, we see a small increase in the star-forming gas mass over the ${\sim}1$~Gyr timescale which is not seen without magnetic fields.
In both simulations, approximately 1/4 of the initial star-forming gas is converted to stars on this timescale.
This gas is replenished through accretion onto the satellite.
The increase in star-forming gas mass suggests that the accretion rate is higher for this satellite in the simulation with magnetic fields.
A lower accretion rate onto the ISM in the simulation without magnetic fields is consistent with stronger ram pressure stripping as more of the satellite's CGM has been removed, therefore diminishing the reservoir from which to replenish gas which has formed stars.

We note that the ram pressure stripping here is less extensive than in galaxy clusters where the more extreme environment strips the satellite's CGM faster, exposing its ISM to ram pressure.
Simulations of ram pressure stripping in cluster-like environments \citep[e.g. ][]{ghosh2024} show effective stripping of the satellite CGM while gravity can sufficiently bind the ISM. 
This protects the ISM from the prevailing wind during this time (seen in observations:~\citealt{merluzzi2024}, and in large-volume cosmological simulations:~\citealt{rohr2024}).
In our simulations, lower infall velocities result in lower ram pressure for satellites in the CGM.
This is sufficient to strip the satellite's CGM on these timescales but not to remove it entirely and expose the ISM to ram pressure stripping, explaining the lack of quenching seen here.

The bottom panel of Fig.~\ref{fig:mass_loss_comparison} shows the retained gas fraction at (approximate) pericentre for all satellites in our sample (with $M_{\rm{total, init}} \ge 3 \times 10^{10}$ M$_\odot$) as a function of their closest approach radius, i.e. the (approximate) pericentric distance from the central galaxy.
Our sample consists of 11 satellite galaxies from simulations with magnetic fields and 10 satellites from simulations without magnetic fields.
While some satellites have clear counterparts in each simulation (particularly the massive satellites), not all do.
Lower mass satellites do not all have counterparts in the sample as their properties sit closer to our selection criteria.
Random variations in the simulations lead to slight differences in the pre-stripping mass and in their orbits.
As such, we do not expect pairs of satellites with identical $R_{\rm{min}}$, with and without magnetic fields. 
There are visual differences in the morphology of the satellite tails (seen in Figs.~\ref{fig:all_haloes_nh-t} and~\ref{fig:all_haloes_z-b}) and in the retained gas mass for the two most massive satellites in our sample (Halo 8: Subhalo 1 and Halo 7: Subhalo 1, both with $M_{\rm{total, init}} > 10^{11}$~M$_\odot$).
However, over our full sample of satellite galaxies, we see no significant differences in the fraction of gas stripped between crossing $R_{\rm{200c}}$ and (approximate) pericentre resulting from the presence of magnetic fields.
Median values of retained gas fraction are slightly lower without magnetic fields which indicates stronger stripping, however, there is significant scatter.

The large scatter in the retained fraction of gas mass over our sample is likely due to a combination of the small sample size and the inherent variations in the properties of the satellite population.
These simulations were not specifically run for this work and due to their significant expense, we were unable to obtain a larger simulation sample at equal resolution.
Within our sample, there is are a large range of initial satellite masses from the lower limit of $M_{\rm{total}}=3\times10^{10}$~M$_\odot$ to beyond $10^{11.5}$~M$_\odot$.
Lower mass satellites are more easily stripped due to their shallower gravitational potentials.
Satellites also orbit on a variety of orbital paths.
Some, like Halo 8: Subhalo 1, have more plunging, radial orbits while others orbit through the CGM on wider, more azimuthal trajectories.
Table~\ref{tab:orbits} summarises the variation in orbital type of satellites in our sample.
To compare satellite orbits, we consider their radial velocities relative to the central galaxy.
Around pericentre, satellite motion is almost entirely azimuthal relative to the central galaxy, so $v_r \approx 0$.
Galaxies on plunging radial orbits, are significantly faster at pericentre and spend less time with low radial velocities than satellites on wider, azimuthal orbital paths.
We define `plunging' orbits as those that spend $\le 3$ simulation outputs (${\sim}300$~Myr) with $|v_r|/|v| < 0.5$.
Satellites that spend more than 3 simulation outputs with $|v_r| / |v| < 0.5$, are classified as having `azimuthal' orbits.
A threshold of $|v_r|/|v| = 0.5$ is a natural choice, because the radial velocity component dominates above this value, whereas the azimuthal velocity is dominant below this threshold.
The classifications are robust to variations in the radial velocity threshold (e.g. $|v_r| / |v| < 0.4$ or $< 0.6$) with no changes to the orbit classifications in Table~\ref{tab:orbits}. 
In addition to significant differences in orbital path, we also see variation in the orientation of the satellites.
Some are oriented `face-on' to the gas of the host's CGM, presenting a much larger surface area than a galaxy moving through the CGM with an `edge-on' orientation. 
Thus, a face-on satellite experiences a greater force from ram pressure.
Given the variations in initial mass and orientation with respect to the ram pressure in our sample, we do not find any difference in retained fraction of gas mass as a result of orbital type.

\begin{table}
    \centering
    \begin{tabular}{ccc}
        \hline
        $B$ fields & Plunging & Azimuthal\\
        \hline
        Yes & 6 & 5  \\
        No  & 6 & 4  \\
        \hline
    \end{tabular}
    \caption{Summary of orbital types of the satellite galaxies in our sample. 
    Orbits are defined as `plunging' if the satellite spends $\le3$ simulation outputs (${\sim}300$~Myr) with $|v_r| / |v| < 0.5$. Satellites on `azimuthal' orbits spend $>3$ simulation outputs with $|v_r| / |v| < 0.5$. For all satellites in our sample, 3 simulation outputs amount to $\le50$~per cent of the time the satellites spend within $R_{\rm{200c}}$. This diversity in orbital path contributes, in part, to the scatter seen in the lower panel of Fig.~\ref{fig:mass_loss_comparison}. However, other factors, including inclination to the CGM flow, also play a role.}
    \label{tab:orbits}
\end{table}

\subsection{Ram pressure stripping of massive satellites: CGM Jellyfish}

When the most massive ($M_{\rm{200c}}>10^{11.5}$~M$_\odot$) subhaloes with the largest gas reservoirs undergo ram pressure stripping, they contain a sufficient quantity of gas to form extended stripped tails.
The tails of these `CGM jellyfish' galaxies are similar to, though lower density than, the tails of jellyfish galaxies in clusters~\citep{goller2023}.
CGM jellyfish tails show significant differences with and without magnetic fields.
Figure~\ref{fig:Satellite_Gas_with_without_B} shows projections of (from left to right) the hydrogen number density, temperature, metallicity and vorticity of Halo~8: Subhalo~1 with (top) and without (bottom) magnetic fields near pericentre ($R_{\rm{min}} = 112$~kpc).
Vorticity here is the curl of the cell velocity vector.
The satellite is rotated such that its velocity vector points in the negative $z$-direction.
We find tails of dense, cool gas visibly extending more than 100 kpc downwind from the satellite.
Without magnetic fields, the tail is longer, extending beyond 200 kpc.
The satellite galaxy's gas disc is much smaller in the simulation without magnetic fields.
The reduced disc size is unlikely to be a result of ram pressure stripping as this has also been seen for Milky Way-mass central galaxies~\citep{whittingham2021}.

Differences in the tails between the two simulations are most pronounced in the distribution of metals.
Without magnetic fields, the region of high metallicity in the wake of the satellite is much broader and longer than with magnetic fields.
While higher metallicities are largely co-spatial with cool gas in the simulation with magnetic fields, this is not the case in the simulation without magnetic fields.
Rather, the metallicity is also higher in large regions containing hotter, lower density gas around the dense tail.
The wider distribution of metals into the hot gas, combined with higher vorticity in the hot gas surrounding the tail, indicates greater turbulence and thus turbulent mixing in the absence of magnetic fields.
As these mixing signatures are less prevalent in simulations with magnetic fields, this indicates a degree of suppression of fluid instabilities (e.g. the Kelvin-Helmholtz instability: \citealt{walker2017}, \citealt{shi2023}) by magnetic fields.


\subsection{Origin and evolution of satellite tail gas}\label{sec:tracer-analysis-results}

To quantify mixing, we determine the origin of the cool, dense gas in our CGM jellyfish tails.
We select all gas cells in the tail based on their location relative to the satellite galaxy and their density.
Our cylindrical selection region extends from 30~kpc to 360~kpc downwind of the satellite with a radius of 150 kpc.
This region is represented in Fig.~\ref{fig:Satellite_Gas_with_without_B} by the dashed white rectangles.
This spatial criterion ensures we do not select gas found in the satellite's ISM, but that we include the entire tail visible in projections of gas density.
To exclude the hot, low-density gas of the host CGM, we place a density criterion of n$_{\rm{H}} \ge 2\times 10^{-4}$~cm$^{-3}$ on gas within the cylindrical region.

We use Monte-Carlo tracer particles (hereafter, tracer particles or tracers) to determine the origin of this gas.
At (approximate) pericentre, we select all tracer particles contained within gas cells in the cylindrical region which meet the density criterion.
We then follow the satellite back in time to where it crosses $R_{\rm{200c}}$.
At each simulation output time, we record the tracer particles associated with the satellite subhalo.
The tracers associated with the subhalo at any output time during this evolution are compared to the tracer particles in the tail.
Any tracer particles in the tail which were previously associated with the subhalo while located within $R_{\rm{200c}}$ of the host halo are categorised as `stripped', while any tracers found in the tail which were not previously associated with the subhalo are categorised as `condensed'.

When applying this to Halo 8: Subhalo 1 and Halo 7: Subhalo 1, we select more than $10^5$ tracer particles from the satellite tails in each simulation.
For both haloes, we select significantly more tracer particles (factor ${\gtrsim}4$) from the tails in the simulations without magnetic fields.
The initial tail mass selected for Halo 8: Subhalo 1 is ${\sim}6.3\times10^{9}$~M$_\odot$ with magnetic fields and ${\sim}2.2\times10^{10}$~M$_\odot$ without magnetic fields.
Of this, approximately 2/3 of cells contain a tracer particle and thus we trace approximately 2/3 of the total tail mass.

\begin{figure}
	\includegraphics[width=\columnwidth]{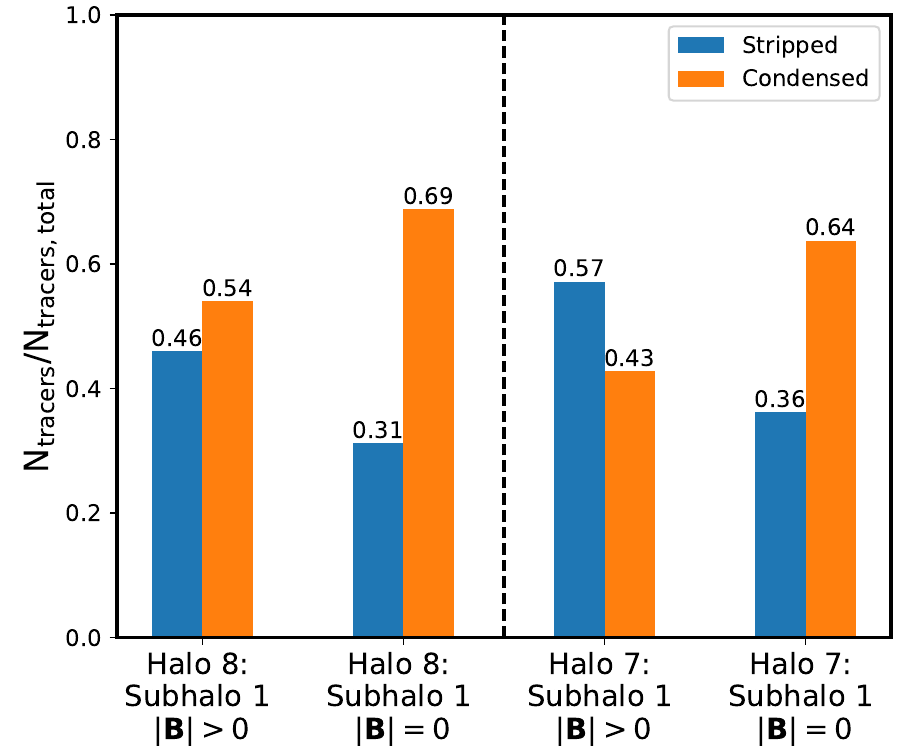}
    \caption{
    Fractions of tracer particles in the tails of two satellite galaxies (Halo~8: Subhalo~1 and Halo~7: Subhalo~1) stripped from each satellite galaxy and its CGM (blue bars) or condensed from the CGM of the host halo (orange bars).
    Fractions are displayed for simulations with (1st and 3rd sets of bars) and without (2nd and 4th sets) magnetic fields.
    In the simulations without magnetic fields, the gas condensed from the host CGM makes up a larger fraction of the gas in the tails than in the simulations with magnetic fields.
    This shows that mixing is more efficient in the absence of magnetic fields.}
    \label{fig:Tail_Tracer_Composition}
\end{figure}

Figure~\ref{fig:Tail_Tracer_Composition} shows the fraction of `stripped' and `condensed' tracer particles found in the tails of two massive satellites, simulated with and without magnetic fields.
Whether magnetic fields are present or not, more than 40\% of tracers particles (and thus gas) within the tails of these galaxies originated in the CGM of the host halo.
For simulations with magnetic fields, the tails consist of approximately equal parts stripped gas and condensed gas.
However, the fraction of condensed gas is considerably larger in simulations without magnetic fields, where condensed tracer particles outnumber stripped tracers by a factor of ${\sim}2$.
In the top panel of Fig.~\ref{fig:mass_loss_comparison}, we saw that the mass loss in the simulation without magnetic fields is approximately double that of the simulation with magnetic fields.
Given the larger quantity of stripped gas and larger condensed fraction of the tail, the amount of condensed gas in the tail is significantly larger (by a factor of ${\gtrsim}4$) in simulations without magnetic fields.
This suggests that there is an elevated rate of mixing in simulations without magnetic fields, more efficiently distributing metals from the stripped gas into the hot CGM of the host as was seen previously in the projections in Fig.~\ref{fig:Satellite_Gas_with_without_B}.
This enhances cooling and therefore condensation onto the stripped gas.

\begin{figure*}
    \centering
    \includegraphics[width=\textwidth]{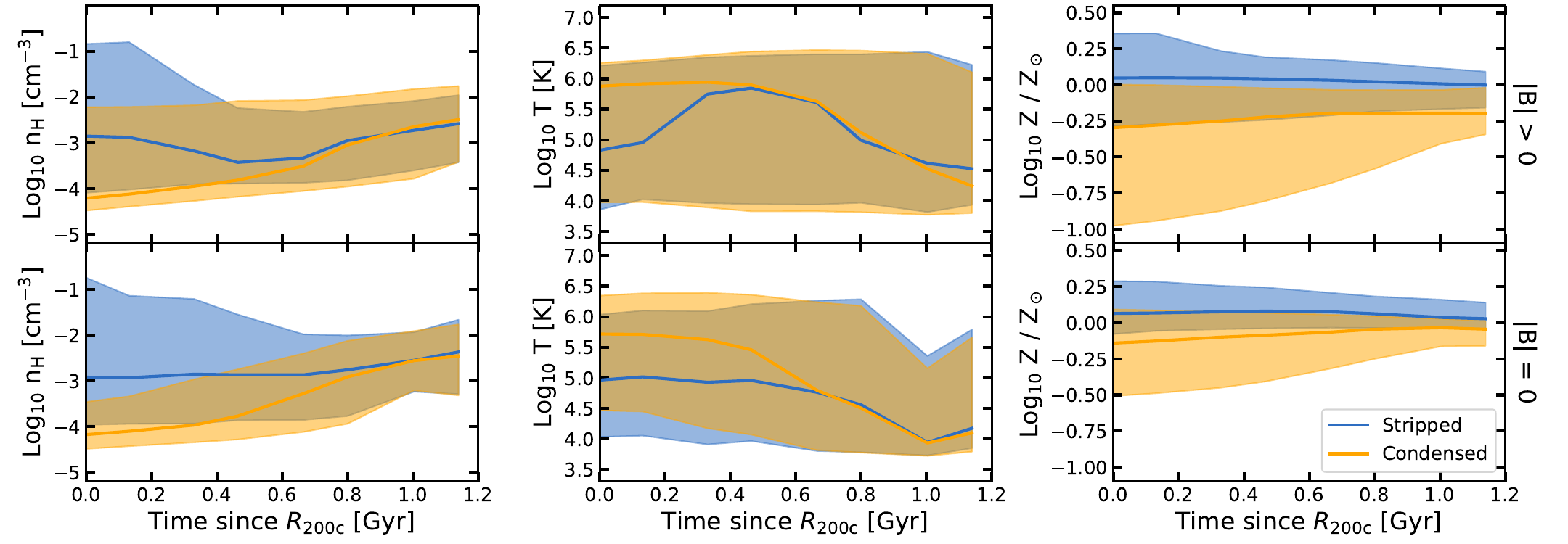}
    \caption{The evolution of the hydrogen number density, temperature and metallicity for the stripped (blue curves) and condensed (orange curves) components of the cool gas in the tail of Halo 8: Subhalo 1 with (top) and without (bottom) magnetic fields. Gas cells were selected at $z=0.01$ when the satellite was close to pericentre. This corresponds to ${\sim}1.15$~Gyr after crossing $R_{\rm{200c}}$ of the host halo. Gas was followed back in time using Monte-Carlo tracer particles to the simulation output closest to when the satellite crossed $R_{\rm{200c}}$. At this time, stripped gas was located within the satellite system and condensed gas in the host CGM. Solid curves denote the median values. Shaded regions show the 16\textsuperscript{th} to 84\textsuperscript{th} percentile range. Both with and without magnetic fields, the condensed gas is initially lower density and hotter than the stripped gas. Given its origin in the host CGM, this is expected. Without magnetic fields, the median metallicity of the condensed gas is higher throughout the evolution compared to the simulation with magnetic fields. This indicates that the CGM environment is more strongly mixed without magnetic fields. While the median metallicity values of the stripped and condensed samples move closer together in both simulations, without magnetic fields, the median values converge further on these timescales. This suggests enhanced mixing between the stripped and condensed populations in the tail in the simulation without magnetic fields.}
    \label{fig:tail_metallicity_mixing}
\end{figure*}

\begin{figure*}
	\includegraphics[width=\textwidth]{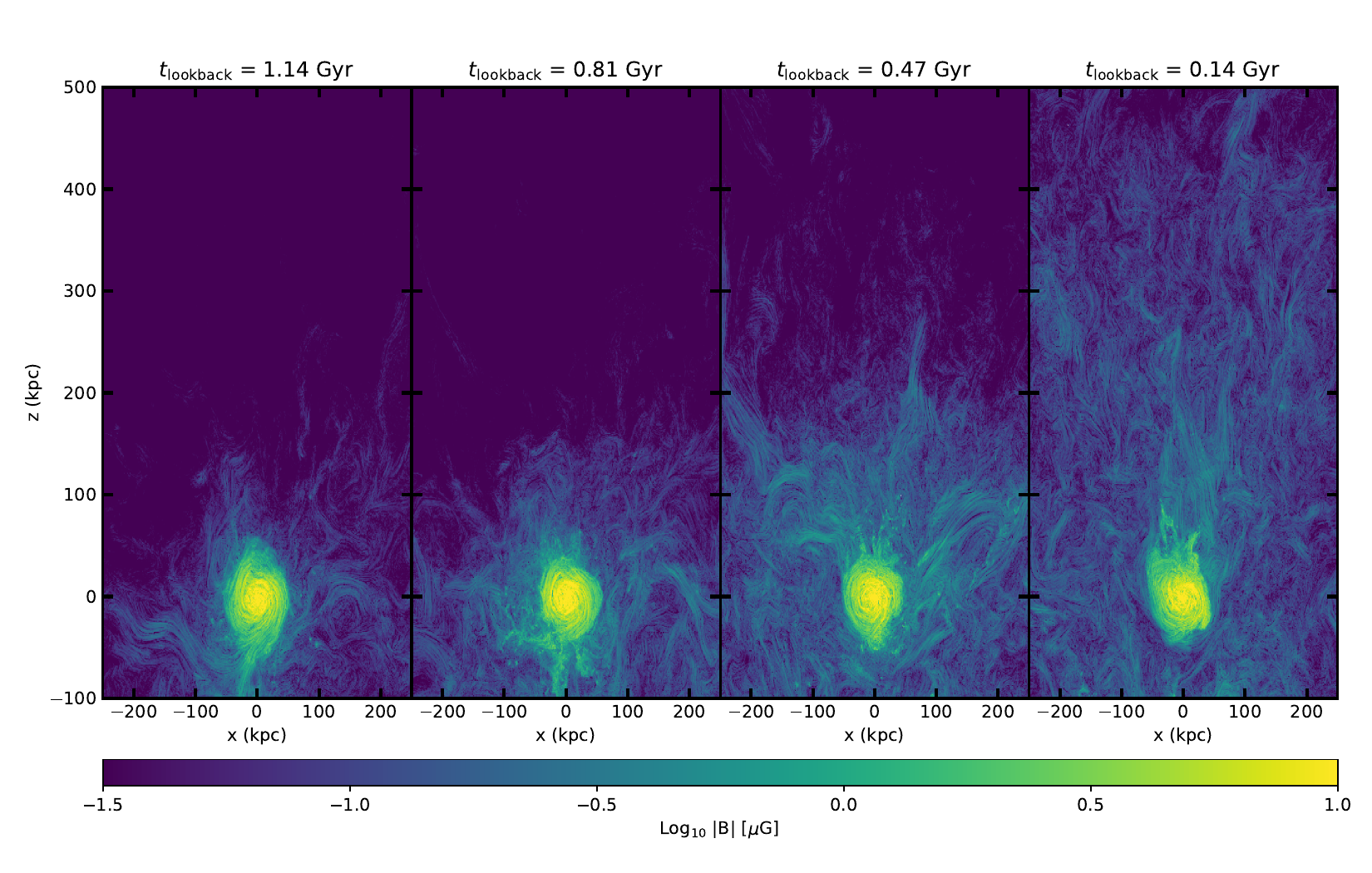}
    \caption{Projections of the magnetic field strength for Halo 8: Subhalo 1. 
    The satellite is oriented as in Fig.~\ref{fig:Satellite_Gas_with_without_B} and is projected at a depth of 50~kpc.
    The magnetic field lines are overlaid using the Line Integral Convolution method \citep{cabral1993} to show the structure of the magnetic field. 
    As the satellite falls into the inner CGM of the host halo, the density of the CGM and therefore the ambient magnetic field strength increases. A magnetic draping layer forms around the leading edge of the satellite with the magnetic field lines forming the characteristic `swept back' structure in the magnetic field (also seen in~\citealt{pfrommer2010} and~\citealt{sparre2024a}). At later times, the magnetic field downwind of the satellite aligns with the stripped tail as seen in observations~\citep{muller2021}. This alignment transverse to the fluid interface likely suppresses mixing between the stripped gas and the host CGM.}
    \label{fig:B_field_evolution}
\end{figure*}

We saw in Fig.~\ref{fig:Satellite_Gas_with_without_B} that there are significant differences between the satellite galaxy tails in simulations with and without magnetic fields, particularly when considering the distribution of metals into the surrounding host CGM.
Using the same tracer selection process as for Fig.~\ref{fig:Tail_Tracer_Composition}, we followed the evolution of the tracer particle host cells backwards in time for the `stripped' and `condensed' populations.
Figure~\ref{fig:tail_metallicity_mixing} shows the evolution in the median value and 16\textsuperscript{th} to 84\textsuperscript{th} percentile ranges of the hydrogen number density, temperature and metallicity for the stripped and condensed tracer populations for Halo 8: Subhalo 1 with and without magnetic fields.
Tracers were selected when the satellite was approximately at pericentre.
For Halo 8: Subhalo 1, this occurred at $z\approx0.01$, corresponding to ${\sim}1$~Gyr after the satellite crossed $R_{\rm{200c}}$ of the host halo.
Using the selected tracer particles, the stripped and condensed populations were followed back through the simulation to when the satellite crossed $R_{\rm{200c}}$.
At this time, the stripped gas was located within the satellite, while the condensed gas was located in the host CGM.

The stripped gas is initially cooler, denser and higher metallicity than the condensed gas.
As the satellite halo is significantly lower mass than the host, the lower temperature of the stripped satellite CGM is unsurprising.
That this gas is also higher in metallicity than the ambient CGM indicates that gas is being stripped from the inner CGM of the satellite which is more metal-rich.
There are no significant differences in the median values of the density or temperature between the simulations with and without magnetic fields.
As expected, the condensed gas sample is initially (i.e. while in the host CGM) hotter and at lower density than the stripped gas which originates in the satellite galaxy.
The initial scatter of the density of the condensed component is smaller without magnetic fields while the median value is similar in both simulations, suggesting that the host CGM is more strongly mixed in the absence of magnetic fields.
The scatter in temperature is similar between the stripped and condensed gas samples in both simulations.
The median temperature of the stripped gas is relatively high (${\sim}10^5$ K) indicating that the majority of stripped gas is from the satellite CGM as opposed to the cooler ISM as shown in Fig.~\ref{fig:mass_loss_comparison}.

With magnetic fields, we initially see a wider spread in metallicity for both the stripped and condensed populations than in the simulation without magnetic fields.
In both cases, the difference between median metallicity values for the selected gas samples decreases over time as the samples become more similar.
We also see the scatter of both samples reduce as this occurs.
Without magnetic fields, the median metallicity of the condensed gas is higher with a smaller scatter than with magnetic fields.
Similarly to the differences in the density scatter, this indicates that the host CGM is generally better mixed in the absence of magnetic fields.
We find that the difference between the median values of the selected samples decreases over time in both simulations, though the decrease is slightly larger without magnetic fields.
The decrease in overall scatter is larger with magnetic fields. 
This is likely due to initial differences in the host halo where gas is generally less mixed with more extremely metal-poor gas.

\subsection{Magnetic draping of satellite galaxies}

The differences in the evolution of these properties (Fig.~\ref{fig:tail_metallicity_mixing}), as well as the stripping rate (Fig.~\ref{fig:mass_loss_comparison}) could be explained by magnetic draping.
Magnetic draping occurs when an object, in this case a satellite galaxy, moves through a magnetised medium such as the host CGM. 
Magnetic field lines curve around the object forming a draping layer with increased magnetic field strength~\citep{dursi2008, pfrommer2010}.
This draping layer can protect the gas in the satellite against ram pressure stripping (e.g.~\citealt{sparre2024}) because the increased magnetic pressure acts to limit vorticity at the interface between fluids (in this case the cooler gas of the satellite and the hot gas of the host CGM) and suppresses the formation of fluid instabilities~\citep{shi2023}.

Figure~\ref{fig:B_field_evolution} shows projections of the magnetic field strength of Halo~8: Subhalo 1 and the surrounding gas at various times spanning ${\sim}1$~Gyr from left to right.
Over this period, the satellite falls inwards towards the central galaxy, moving deeper into the host halo.
The satellite is rotated as in previous figures with its velocity vector pointing in the negative $z$-direction.
The structure of the magnetic field is overlaid as brushstrokes using the Line Integral Convolution method~\citep{cabral1993}.
As the satellite falls towards the central galaxy, the ambient magnetic field strength increases.
Ram pressure becomes stronger as the host CGM density rises and as a result, more gas is stripped from the satellite.
We see in the 3rd and 4th panels that the leading edge of the satellite is swept back towards its centre and the downwind edge of the disc becomes elongated along the $z$-direction.
The magnetic field lines ahead of the satellite become more perpendicular to the velocity vector of the satellite and form the characteristic swept back morphology of magnetic draping also seen in~\cite{dursi2008} and~\cite{sparre2024}.
The draping layer likely suppresses fluid instabilities at the leading edge of the galaxy, reducing the rate of ram pressure stripping.

In the final panel of Figure~\ref{fig:B_field_evolution}, we see the magnetic field lines downwind of the satellite align with its tail, preferentially oriented along the $z$-axis.
At earlier times, they displayed no coherent structure.
This alignment of the magnetic field with the tail has also been observed in clusters~\citep{muller2021} and found in ICM wind-tunnel simulations~\citep{sparre2024}.
The magnetic field alignment transverse to the fluid interface likely suppresses fluid instabilities similar to the effect of the draping layer at the leading edge of the satellite.
This suppression results in the reduced mixing and subsequent condensation seen in simulations with magnetic fields.

\section{The impact of CGM refinement}\label{sec:CGM-Refinement}

\begin{figure}
	\includegraphics[width=\columnwidth]{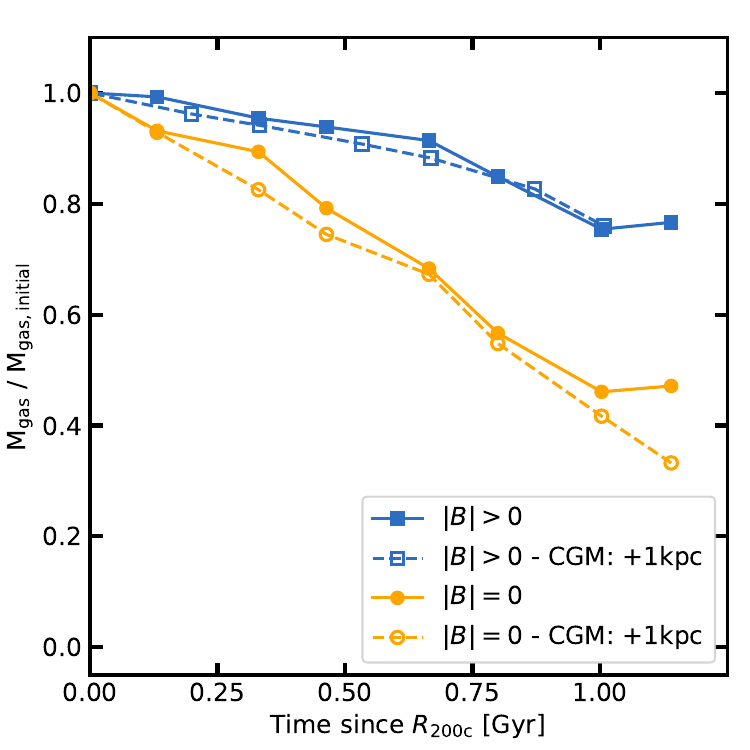}
    \caption{Comparison of the gas mass evolution for Halo 8: Subhalo 1 in 4 simulations of this halo. Blue lines, marked with squares, show simulations with magnetic fields. Orange lines, marked with circles, show simulations without magnetic fields. Dashed lines with unfilled markers denote simulations with 1 kpc spatial refinement in the CGM. Solid lines with filled markers denote simulations at the standard mass resolution described in Table~\ref{tab:halo_table}. The fraction of gas mass removed from this satellite is nearly identical in the standard simulations and their counterparts with CGM refinement. This shows that the mass loss is converged, indicating that any differences between simulations with and without magnetic fields are due to physical processes - i.e. the inclusion of magnetic fields.}
    \label{fig:CGM_stripping_comparison}
\end{figure}

\begin{figure}
	\includegraphics[width=\columnwidth]{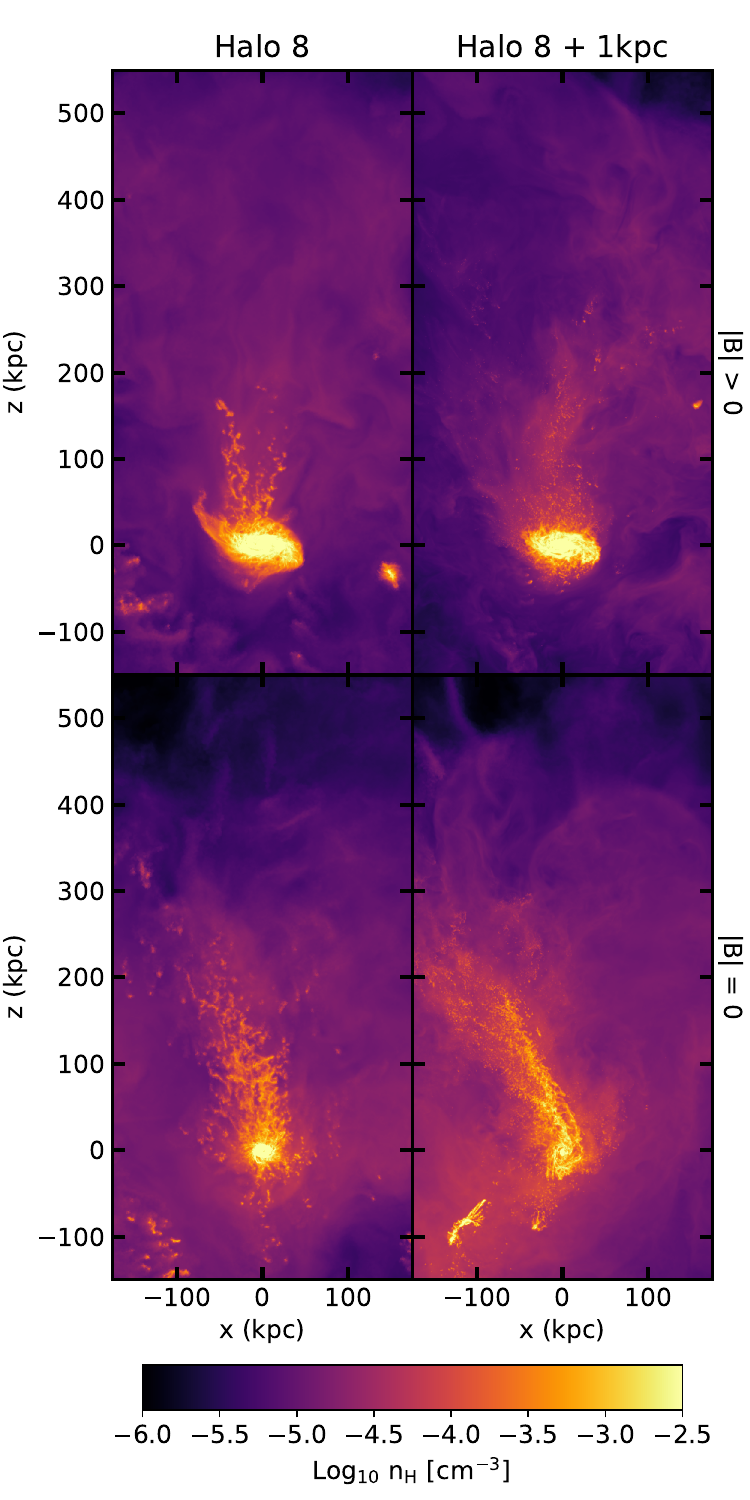}
    \caption{Projections of the hydrogen number density for Halo 8: Subhalo 1 in the 4 simulations of Halo 8. Each simulation has been rotated such that the velocity vector of the satellite points in the negative $z$-direction. The satellite is projected with a projection depth of 50 kpc.
    The simulations shown in the upper panels include magnetic fields, whereas those in the lower panels do not. The left column shows the standard mass resolution simulations. The right column includes a 1 kpc spatial refinement in the CGM in addition to mass refinement.
    As in the standard simulations, the tail with magnetic fields displays smaller clumps than in the simulation without magnetic fields.
    In both cases with CGM refinement, the clumps are smaller than in the standard resolution simulations which is expected given the improved spatial resolution~\citep{ramesh2024}.
    Smaller, but more numerous clumps cause the tails to appear more visibly diffuse. 
    }
    \label{fig:CGM-Tail-Morph}
\end{figure}

To check that our results are converged, we use our simulations of Halo 8 with 1~kpc spatial refinement.
As above, we focus our analysis on Halo 8: Subhalo 1 and thus make comments on convergence for massive satellites only.
We leave a detailed study of the lower mass satellites for future work with higher resolution simulations. However, for this single halo simulated at higher resolution, we see considerable scatter in retained gas fraction as in Fig.~\ref{fig:mass_loss_comparison}.

Figure~\ref{fig:CGM_stripping_comparison} shows the retained gas fraction of the entire subhalo over time for the standard resolution and CGM refinement simulations of Halo 8: Subhalo 1, with and without magnetic fields.
The initial gas mass and the start of stripping are defined at the point when the satellite crossed $R_{\rm{200c}}$.
Both with and without magnetic fields, we see that the evolution of the satellite's total gas mass is nearly identical between the standard and CGM refinement simulations.
This shows that ram pressure results in similar stripping even when the host and satellite CGM is simulated with ${\sim}39$~times the number of resolution elements. 
The same is true for star-forming gas mass (not shown).

Because the mass loss rates are converged with resolution, we now explore the effects of CGM refinement on the tails of CGM jellyfish.
Figure~\ref{fig:CGM-Tail-Morph} shows projections of hydrogen number density for Halo 8: Subhalo 1 in the simulations at standard resolution and those with CGM refinement, with and without magnetic fields. 
The satellite is rotated as in previous figures.
While the overall mass loss from the satellites is unaffected by the inclusion of CGM refinement, there are significant differences in the morphology and extent of the tails.
With CGM refinement, the tails appear more diffuse in density projections than their mass refined counterparts, whether magnetic fields are included or not.
However, the median density of the cool gas is slightly higher with CGM refinement, consistent with other applications of this refinement scheme which show increased cool gas column densities~\citep{vandevoort2019}.

In the simulation with both magnetic fields and CGM refinement, the tail morphology is particularly different with much smaller clumps in the tail.
This more diffuse tail extends further behind the satellite than in the standard case to beyond 200 kpc downwind.
Without magnetic fields, the tail is approximately the same length with CGM refinement as it is at standard resolution. However, clumps in the tail are smaller and more numerous as also seen in the simulation with CGM refinement and magnetic fields.
More numerous but smaller clouds at higher resolutions have also been seen in other works employing additional refinement schemes (e.g. the mass-refinement version of CGM refinement presented by ~\citealt{ramesh2024}).
Despite their more diffuse appearance, we see the same evolutionary behaviour in density, temperature and metallicity with CGM refinement as was seen at standard resolution (not shown, see Fig.~\ref{fig:tail_metallicity_mixing} for standard resolution).
For both resolutions, the stripped gas is initially denser, cooler and higher metallicity than the condensed gas.
Additionally, without magnetic fields, the difference in the median metallicity of the samples decreases more over time than in the simulation with magnetic fields.
These conclusions are therefore converged with resolution.



\section{Discussion and Conclusions}\label{sec:disc_and_conc}

We examined the effect of magnetic fields on the ram pressure stripping of satellite galaxies within the CGM of massive galaxies in cosmological, magnetohydrodynamical simulations using the moving mesh code \textsc{arepo} as part of the SURGE project.
We simulated three massive haloes with and without magnetic fields.
One of these haloes was also simulated with enhanced spatial resolution in the CGM.
The main conclusions of this work are as follows:

\begin{enumerate}
    \item There are modest differences in the host CGM environment in simulations with and without magnetic fields (Figs.~\ref{fig:all_haloes_nh-t} and~\ref{fig:all_haloes_z-b}).
    Without magnetic fields, the density of the host CGM is slightly higher, particularly in central regions.
    The temperature in the outer CGM is slightly lower without magnetic fields, and the distribution of metals throughout the halo is smoother, indicating the host CGM is somewhat better mixed.
    With magnetic fields, there is a larger difference between the regions of highest and lowest metallicity.
    \item We see no clear difference in the fraction of gas mass retained by satellites undergoing ram pressure stripping between simulations with and without magnetic fields over our full sample (Fig.~\ref{fig:mass_loss_comparison}). However, because of the small size of our sample and the inherent variation in the physical and orbital properties of the satellite population, there is significant scatter within the population. We find that the median retained gas fraction is slightly lower without magnetic fields though this difference is well within the scatter range. Increasing the number of satellites studied is needed to determine whether this is a significant result. 
    \item For the two most massive satellites in our sample, significantly more gas is stripped in the absence of magnetic fields. This gas is primarily stripped from the CGM of the satellite (Fig.~\ref{fig:mass_loss_comparison}). This stripped gas forms long tails in the wake of the galaxy similar in appearance to, though much lower density than, the tails of jellyfish galaxies found in clusters. We refer to these satellites as `CGM jellyfish'. 
    \item CGM jellyfish galaxies show significant differences in the extent, mass and morphology of their stripped tails (Fig.~\ref{fig:Satellite_Gas_with_without_B}). Without magnetic fields, tails are visibly longer in density projections, with more cool gas in the wake of the satellite. These longer tails are formed of larger clumps than in simulations with magnetic fields. The greatest difference in the tails is seen in metallicity. In simulations with magnetic fields, high metallicity regions are generally co-spatial with the cool gas tail. Without magnetic fields, high metallicity regions span a larger volume than the dense tail. Metals have been distributed more widely into the local environment through turbulent mixing. Increased turbulent mixing is also indicated by higher vorticity in simulations without magnetic fields.
    \item Examining the origin of gas in the tails of CGM jellyfish shows substantial mixing between stripped gas and the host CGM environment both with and without magnetic fields (Figs.~\ref{fig:Tail_Tracer_Composition} \& \ref{fig:tail_metallicity_mixing}). 
    For the satellite examined in detail in this work, the tail consists of cool gas with mass on the order $10^{10}$~M$_\odot$.
    As expected from the enhanced stripping rate, without magnetic fields, the tail is more massive by a factor ${\sim}2$.
    When magnetic fields are included, approximately half of all cool gas in the tail does not originate from the satellite's ISM or CGM.
    Rather it has condensed from the ambient environment. 
    Without magnetic fields, this condensed fraction constitutes approximately 2/3 of the gas in the tail because of stronger turbulent mixing and induced cooling. When coupled with the enhanced stripping, the mass of condensed gas is at least a factor of 4 higher in the absence of magnetic fields.
    \item The difference in ram pressure stripping with and without magnetic fields is likely due to draping of the magnetic field around the satellites as they move through the magnetised host environment (Fig.~\ref{fig:B_field_evolution}). The draping layer protects the leading edge of the satellite, suppressing fluid instabilities and reducing the overall stripping due to ram pressure. Magnetic fields are found to align with the tail of the CGM jellyfish galaxy, similar to that observed in clusters by~\cite{muller2021}. This alignment of the magnetic field along the tail suppresses fluid instabilities between the cool gas of the tail and the hot gas of the environment, reducing turbulent mixing and the associated induced condensation.
    \item We repeated our simulations for one halo with an additional spatial refinement criterion in the CGM and found very similar results in retained gas fraction and the effect on the CGM jellyfish tail with and without magnetic fields (Figs.~\ref{fig:CGM_stripping_comparison} \& \ref{fig:CGM-Tail-Morph}). We conclude, therefore, that our results are robust to changes in resolution. The most notable change due to resolution is in the structure of the tails themselves. With CGM refinement, the clouds that form the satellite tails are smaller and more numerous than at lower resolution. As a result, tails in simulations with CGM refinement appear more diffuse than at standard resolution even though the gas densities tend to be slightly higher.
\end{enumerate}

The topic of ram pressure stripping of satellites has also been explored in large-volume cosmological simulations.
Exploring its effects in the TNG50 simulation, \cite{rohr2023a} found that satellite galaxies in $10^{13}$~M$_\odot$ haloes tend to undergo stripping for approximately 5~Gyr before the mass of the ISM dropped below their resolution limit (i.e. the satellite is quenched). 
In this time, satellites typically complete a full orbit. We have only explored galaxies on initial infall (timescales of ${\sim}1$~Gyr).
This timescale is identified as the period of peak ram pressure stripping by \cite{rohr2023a}.
Satellites in TNG50 show similar stripping rates as the satellites in our simulations with magnetic fields.

\cite{sparre2024} conducted idealised wind-tunnel simulations for a Milky-Way mass galaxy (${\gtrsim}0.5$ dex more massive than the satellites in this work) in a time-evolving cluster-like wind.
This wind is hotter, denser and faster than the host CGM wind experienced by our satellites.
They also used \textsc{arepo} and the Auriga model, but at a lower resolution 
than our simulations.
In agreement with our work, \cite{sparre2024} find that without magnetic fields, ram pressure stripping removes significantly more gas from the galaxy.
They find that the magnetic field aligns with the tail as seen in our simulations and shown in Fig.~\ref{fig:B_field_evolution}.
They also find that increasing the resolution leads to smaller but more numerous clouds in the tail.
The size of these clouds is limited by resolution.
This lack of convergence is also seen in our simulations (Fig.~\ref{fig:CGM-Tail-Morph}) and in the mass-refined CGM refinement simulations of \cite{ramesh2024}.

We have demonstrated that magnetic fields can significantly affect the ram pressure stripping rates of satellite galaxies in the CGM around massive galaxies and the subsequent evolution of their tails.
We see consistent results for the most massive satellites in our sample, in different haloes and independent of resolution.
However, we wish to stress that, due to constraints on computation time and the significant computational cost of these simulations, we have only studied a small sample of satellite galaxies in a small number of haloes between $M_{\rm{halo}} = 10^{12.5}$~M$_\odot$ and $M_{\rm{halo}} = 10^{13}$~M$_\odot$ at low redshift ($z < 0.5$).
Further work is needed to determine the prevalence of these effects on the wider satellite galaxy population, particularly at lower mass, and at higher redshift. 

Observationally exploring ram pressure stripping of satellite galaxies around massive galaxies is challenging.
On the timescales examined in this work, we find that ram pressure strips the CGM of the satellites.
Direct observational signatures are, therefore, difficult to detect because the density of the stripped gas is significantly lower than for cluster jellyfish galaxies.
However, the magnetic field structure of the draping layer, indicating suppression of ram pressure stripping may be observable.
This would require polarised background sources enabling rotation measure observations.
Due to its large angular size, this is likely most practical to test with observations of the Large Magellanic Cloud, which is currently being stripped by the Milky Way CGM.

Although our selection criteria were chosen to ensure well resolved satellites with saturated magnetic fields, any structures below the kpc-scale regime (e.g. shocks or turbulence) cannot be resolved.
This includes the structure of the tail itself which as is seen with the CGM refinement simulations, is not yet converged.
When combined with the temperature floor of our simulations, this lack of convergence makes any examination of the long-term evolution of the stripped cool gas challenging and is therefore left to future work.
It is also possible that our simulations underestimate the amplification of the magnetic fields in these satellites.
Additionally, \cite{werhahn2025} have shown that satellite galaxies have stronger magnetic fields after passing pericentre.
We conclude that examining a broader sample of satellite galaxies over longer periods can provide a more comprehensive view of the wider role of magnetic fields in ram pressure stripping of satellites in the CGM of massive galaxies and in galaxy groups.



\section*{Acknowledgements}

We thank the referee for their useful comments that improved the paper.
We would like to thank Christoph Pfrommer for useful discussions which aided this study.
TAR is supported by the \href{https://cdt-aimlac.org/}{UKRI AIMLAC CDT}, funded by grant EP/S023992/1. 
FvdV is supported by a Royal Society University Research Fellowship (URF\textbackslash R1\textbackslash 191703 and URF\textbackslash R\textbackslash 241005).
ATH is supported and funded by the \href{https://data-intensive-cdt.ac.uk/}{Data Intensive CDT}, on behalf of the Science and Technology Facilities Council (STFC), as part of grant ST/P006779/1.
RB is supported by the SNSF through the Ambizione Grant PZ00P2\_223532.
This work used the DiRAC@Durham facility (under project code dp221) managed by the Institute for Computational Cosmology on behalf of the STFC DiRAC HPC Facility (www.dirac.ac.uk). 
The equipment was funded by BEIS capital funding via STFC capital grants ST/K00042X/1, ST/P002293/1, ST/R002371/1, and ST/S002502/1, Durham University and STFC operations grant ST/R000832/1. 
DiRAC is part of the National e-Infrastructure.
The authors gratefully acknowledge the Gauss Centre for Supercomputing e.V.\ (https://www.gauss-centre.eu) for funding this project (with project code pn68ju) by providing computing time on the GCS Supercomputer SuperMUC-NG at Leibniz Supercomputing Centre (https://www.lrz.de).

\section*{Data Availability}

The data underlying this article will be shared on reasonable request to the corresponding author.



\bibliographystyle{mnras}
\bibliography{Paper_Bibliography_RPS_Satellites_B_noB.bib} 





\bsp	
\label{lastpage}
\end{document}